# Ultimate Depth Resolution and Profile Reconstruction in Sputter Profiling with AES and SIMS

Siegfried Hofmann

Max-Planck-Institut fuer Metallforschung
Seestr. 92, D-70174 Stuttgart, Germany
e-mail: hofmann@mf.mpi-stuttgart.mpg.de

New developments in quantitative sputter depth profiling during the past ten years are reviewed, with special emphasis on the experimental achievement of ultrahigh depth resolution (below 2 nm for sputtered depths larger than 10 nm). In this region, the depth resolution function generally is asymmetric (i.e.non-Gaussian) and it is governed by the three fundamental parameters: atomic mixing length, roughness and information depth. The so called mxing-roughness-information depth (MRI)-model and its application to the quantitative reconstruction of the in-depth distribution of composition, with a typical accuracy of one monolayer and better, is demonstrated for SIMS and AES depth profiles. Based on the combination of the above three parameters, the ultimate depth resolution is predicted to be about 0.7-1.0 nm. Up to now, values of 1.4-1.6 nm have been reported, and the use of low energy molecular ions, e.g. by using sulfur hexafluoride as sputtering gas, has recently pushed the mixing length down to 0.4-0.6 nm. However, particularly in depth profiling of multilayers, it can be shown that minimizing the rouhgness parameter, including the straggling of the mixing length, is most important for the achievement of the ultimate depth resolution.

## 1. Introduction

The most frequently and routinely applied method to obtain the distribution of elemental composition of thin films with depth is ion sputtering in combination with any of the popular surface analysis techniques such as AES, XPS and SIMS (for review, see e.g. [1]). Although precision and accuracy of depth profiling depend on many experimental factors and are difficult to describe comprehensively, a useful characteristic figure of merit is the depth resolution which basically represents the broadening of the measured profile with respect to the original concentration -depth distribution.

Since the beginning of the first systematic studies of the depth resolution (DR) [2-5] and its dependence on various parameters in the mid seventies, the aim is to optimize the DR, i.e. to minimize profile broadening. The following definition of the DR ($\Delta z$) was introduced [3-5] and later adopted by IUPAC and the ASTM E-42 committee: Depth resolution is the depth range over which a signal increases (or decreases) by a specified amount when profiling through an ideally sharp interface between two media. By convention, the „depth resolution corresponds to the distance over which a 16% to 84% (or 84% to 16%) change in signal is measured"[6]. If the resulting shape of such an interface profile can be approximated by an error function, this definition means $\Delta z = 2\sigma$ where $\sigma$ is the standard deviation of the corresponding Gaussian resolution function [1,3-5] which is the derivative of the (measured) error function. Careful separation of the different causes of profile broadening contributions to $\Delta z$ [7] which were shown to add up in quadrature [8] by many experimental and theoretical studies led to a consistent interpretation of depth profiling data as well as to improved measured depth profiles, particularly after the introduction of sample rotation during profiling by A. Zalar in 1985 [9]. It was early recognized that low ion energy (<1keV) and high ion incidence angle (>70°) are essential to optimize $\Delta z$ in case of smooth sample surfaces [5,7]. Today, ion guns capable of sufficiently high sputtering rates and beam quality even at 200 eV energy are available, and the achievable depth resolution is typically about 2 nm and below [1,10-12].

## 2. From Depth Resolution to Depth Resolution Function

While the depth resolution (DR) concept is a simple and useful description of the average composition in a depth range within which nothing is known about the shape

| Phenomenological (Gaussian DRF): | | Shape of Sputter Depth Profiles: | | | |
|---|---|---|---|---|---|
| • Depth Resolution Δz (Hofmann, Ho 1976)[4,5] | | • Diffusional Mixing: Complete in Mixing Range | • Monte Carlo Calculation | • Ballistic Relocation (Boltzmann Transport) | • Diffusion-like Variables |
| *Roughening* (Shimizu 1976 [18], Hofmann, Sanz et al. 1980/81[7,19,20] Carter et al. 1982[21]) | Mixing (Diffusion-like) (Haff & Switkowski 1977 [22], Andersen 1979 [13]) | (Liau et al. 1979 [25]) | (Shimizu et al. 1977-[30]) | (Sigmund et al.1980, Littmark & Hofer 1980) TRIM ( Biersack et al., 1985-[33])-based models [34,35]) | (Collins, Carter et al. 1981 [37]) |
| → Crystalline Orientation (Pamler et al. 1990 [23]) | → Thermochemistry (Cirlin, Cheng et al. 1990 [24]) | → extension to roughening and information depth: (MRI: Hofmann et al.1990-[26,27]) → empirical DRFs: (Zalm et al. 1993- [28], Dowsett et al. 1994-[29]) | | → IMPETUS Program (Armour et al. 1988 [36]) | ( King & Tsong 1984 [38]) |

**Fig. 1**: Schematic survey of the development of phenomenological and atomistic modeling of the depth resolution and the depth resolution function (DRF) in sputter

depth profiling. Note that this is only a limited selection of those researchers and papers which to the author's opinion have been most influencial.

of the original depth distribution, any further data evaluation needs deeper insight in the fundamental mechanisms in sputter profiling in order to find the depth resolution function (DRF) responsible for the deviation of the shape of the measured profile. To achieve this, during the past three decades theoretical models for the parameters that determine DR and the DRF were developed in conjunction with experimental improvement. A number of both phenomenological and atomistic theories were established, , with increasing sophistication with time as schematically shown in **Fig. 1** [1] . The proposed models have at least a twofold purpose: On the one hand they serve as a guideline for optimizing experimental profiling methods, on the other they allow an analytical modeling of the DRF which enables deconvolution of profiles.

The first DRFs for description of profile broadening [13] and deconvolution of profiles [4,14] were simple Gaussian functions. One of the first model predictions of a DRF for practical use in sputter profiling was the so-called SLS (Sequential Layer Sputtering) model based on the increase in surface roughness with sputtered depth according to a Poisson function [1,5]. Although it has been questioned for overestimating roughness and at the same time neglecting atomic mixing, the underlying mathematics for both mechanisms are practically the same for the beginning of sputtering near the surface, i.e. the transient state before sputter equilibrium is established. The SLS model was found very useful for profile reconstruction of the first 10-20 monolayers, sufficient for the quantitative evaluation of native oxide layers or of passive layers in corrosion [7,15]. Recently, a duplex SLS type formalism (for channeling and non-channeling ion sputtering) was found to describe the roughness development with sputtered depth of an Al evaporation layer[16].

For sputter profiling of thicker metallic layers, roughness generally is the dominant contribution to $\Delta z$ [8], represented by a symmetric Gaussian DRF with a single parameter ($\Delta z = 2\sigma$). Particularly since the introduction of sample rotation during profiling [9], sputtering induced roughening of more than a few nm and its increase with sputtered depth can be avoided. High resolution depth profiling with a depth resolution Δz < 5 nm is generally achieved when low energy ions (≤1keV Ar+) are used [1]. Below about Δz=5 nm , atomic mixing and information depth are of increasing importance, and the DRF gets more asymmetric. Ultra-high depth resolution with Δz < 2 nm (7-8 atomic monolayers) is expected for the case of very low ion energy and/or high incidence angle, and by using low energy energy Auger electrons (<100 eV) in AES depth profiling. For ultra-high depth resolution so far only a few experimental examples exist [10, 39, 40]. The ultimate limit seems to be about 3-4 monolayers (Δz=0.7-1.0 nm, [41]). Features smaller than this Δz, e.g. of the order of an atomic monolayer (ML), can only be resolved by deconvolution or profile reconstruction procedures with an appropriate depth resolution function. Accurate experimental determination and theoretical modeling of the depth resolution function are necessary for a correct transformation of the measured sputter profile into the original depth distribution of composition. Such a high accuracy was recently achieved by introduction of the so called MRI-model, for atomic mixing (M), surface roughness (R) and information depth (I) for the description of the depth resolution function.

The MRI model was developed by the end of the eighties in the Applied Surface and Interface Analysis group at the Max-Planck-Institut fuer Metallforschung in Stuttgart [27,42-43] and is described in detail in ref. [11]. A program in Q-Basic has been available since that time and was distributed to some researchers on request. Recently, an improved and more user-friendly version in Visual Basic was developed by Schubert and Hofmann [44] at the National Research Institute for Metals in Tsukuba, and implemented by Yoshihara [45] in the COMPRO software of the Surface Analysis Society of Japan. It is hoped that many fellow researchers use it for quantification of their measured profiles and communication of experience in practical use is encouraged.

In this paper, the basic features of the MRI-model and some applications to AES and SIMS depth profiles will be summarized and discussed in order to demonstrate the usefulness of the approach as well as its capabilities and limitations.

## 3. Deconvolution and Profile Reconstruction (a framework of profile quantification)

Usually, quantitative evaluation of measured profiles is done in the following three steps /1/:

(1) Conversion of the elemental signal intensity into elemental concentration

(2) Conversion of the sputtering time into sputtered depth

(3) Assessment of the depth resolution and of the depth resolution function (theoretically or by measurement) and reconstruction of the original elemental depth distribution.

Tasks (1) and (2) are most important and generate the framework of any quantification. Quantification of SIMS is only straightforward for tracer profiles, otherwise we have to expect strong matrix effects. The latter are less important and fairly well known in quantitative AES and XPS [46], and their use in sputter profiling mainly requires taking into account the near surface composition (within $< 5\lambda$ with $\lambda$ the mean electron escape depth) [47]. Quantification of the depth scale not only requires one point for the sputtering time in the measured profile to be attributed to a certain depth (for example by a stylus measurement of depth the sputtered crater after profiling), but generally has to consider nonlinear time /depth dependencies in case of a composition dependent sputtering rate. For reference samples of Ni/Cr multilayers it was shown that assuming a linear relation with the composition is the key to obtain a correct depth scale [7]. If carried out properly, tasks (1) and (2) establish the linear conditions necessary for step (3). In favourable cases, the time/depth and intensity/concentration relations are already practically linear and can easily be obtained by appropriate sputtering rates and elemental sensitivity factors, respectively. In high resolution depth profiling, a known resolution function applied to a sharp interface can be used to establish the correct relations time/depth and intensity/concentration by fitting of the measured profile with the result of a convolution, as shown below.

Theoretically speaking, sputter depth profiling is the transformation of a real world compositional distribution into an image of it, namely the measured depth profile. This transformation is described by the convolution integral which is governed by the depth resolution function (DRF) g(z-z') [1]:

$$I(z)/I(0) = \int_{-\infty}^{+\infty} X(z') \cdot g(z-z')dz' \tag{1}$$

where I(z)/I(0) is the measured and normalized intensity at the sputtered depth z and X(z') the mole fraction of the respective element at depth z . Deconvolution means solving eqn (1) for X(z') which is possible e.g. by inverse Fourier transformation schemes if g(z-z') and I(z)/I(0) are known. However, such a "reverse" problem often faces practical difficulties because of insufficient data precision too high signal-to-noise level (see discussions of the noise problem by Zalm[48] and Dowsett [49]. Therefore it has become customary to solve the problem by "forward calculation" of the convolution eqn. (1), i.e assuming a suitable X(z') and comparing the calculated profile with the measured profile I(z)/I(0). By changing the input X(z') until an optimum fit is obtained, the "original" in-depth distribution of composition is finally reconstructed [1,11].

## 4. Model calculations and experimental determination of the depth resolution function

### 4.1 Outline of the MRI- model

The beginning of a sputtering profile is characterized by a variation of the DRF with sputtered depth which was early recognized [5]. After typically several monolayers of sputtered depth, sputter equilibrium is obtained (strictly speaking only for homogeneous alloys) and the DRF is approximately constant [1]. The DRF contains the physical parameters determining the "response" of the system under study in terms of the measured profile. These parameters can be divided in the following three categories: change of surface composition (atomic mixing, preferential sputtering etc.), change of surface topography (roughness) and information depth of the analysis method. Although more detailed and complicated atomistic models exist (see Fig. 1 and references in [1]) [36], the so called MRI model based on these three parameters (Mixing-Roughness-Information depth) was recently shown to be able to consistently describe AES depth profiles in GaAs/AlAs multilayers [11,44] as well as SIMS [39,44] profiles.

The MRI- model is capable of giving a mathematical description of the depth resolution function g(z-z'), based on the three fundamental contributions atomic mixing, surface roughness and information depth. Atomic mixing is described by an exponential function with a characteristic mixing zone length, w, the information depth by another exponential function with a characteristic length λ, and the roughness by a Gaussian term with standard deviation σ (corresponding to rms roughness). These functions are employed sequentially to the (assumed) depth distribution of an element, given by thin layers each with (different) concentrations. For example, each monoatomic layer at a location $z_0$ gives a normalized contribution at a sputtered depth z which is described by

atomic mixing:

$$g_w = \frac{1}{w}\exp[-(z-z_0+w)/w] \quad (2a)$$

information depth:

$$g_\lambda = \frac{1}{\lambda}\exp[-(z-z_0)/\lambda] \quad (2b)$$

surface roughness:

$$g_\sigma = \frac{1}{\sqrt{2\pi}\sigma}\cdot\exp\left[\frac{-(z-z_0)^2}{2\sigma^2}\right] \quad (2c)$$

Eqns. (2a-c) can be applied by summing up all the contributions for each depth z after eq. (1), thus representing the calculated depth profile which can be compared to the measured one. This is shown in the examples below (see Fig. 2). A more detailed derivation of the MRI model is given in refs. [11, 50]. Note that in contrast to other "empirical" DRFs consisting of a double-exponential and a symmetric (Gaussian) function used in SIMS [29,51], the MRI parameters have a well defined physical meaning. Therefore they can be theoretically predicted and/or experimentally measured by independent methods. For example, the information depth parameter $\lambda$, is given by the secondary ion escape depth in SIMS (about 1-2 ML) and by the attenuation length of the respective Auger-or photo electrons in AES and XPS, as predicted by Tanuma, Powell and Jablonski [52,53]. The mixing length is at least approximately predicted by the TRIM code [33] (ion ranges or better mean range of total recoil displacements) and can be independently measured by angle dependent AES or XPS [54], as shown below. Roughness is hard to predict, but surface roughness after profiling can be measured by AFM, and original interface roughness by grazing incidence X-ray reflectometry (GIXR). However, as pointed out in ref [11], the "straggling" of the mixing length causes an additional roughness term which is difficult to determine.

4.2 **Experimental Determination of the Depth Resolution Function**

According to eqn. (1), the depth profile of a very thin layer (d→0), a so- called "delta layer" with X(z) = 0 in its vicinity, directly gives I(z) = g(z-z'). Because the maximum normalized signal intensity of a single layer with thickness d decreases approximately with $\Delta z/d$, this method for experimental determination of the DRF is particularly useful in SIMS with its typically high detection sensitivity and strong matrix effects, but can also be used in high resolution AES profiling. In AES, the step function approach generally is more useful. Eqn. (1) shows that for a step function distribution (X(z) = 1 or 0) the DRF g(z-z') is determined by

$$g(z-z') = |dI(z)/dz| \qquad (3)$$

Eqn. (3) means that the absolute value of the differentiated, measured (and normalized) profile I(z) gives the DRF. The depth z can be replaced by the sputtering time t if the sputtering rate $\dot{z} = dz/dt$ is known and constant.

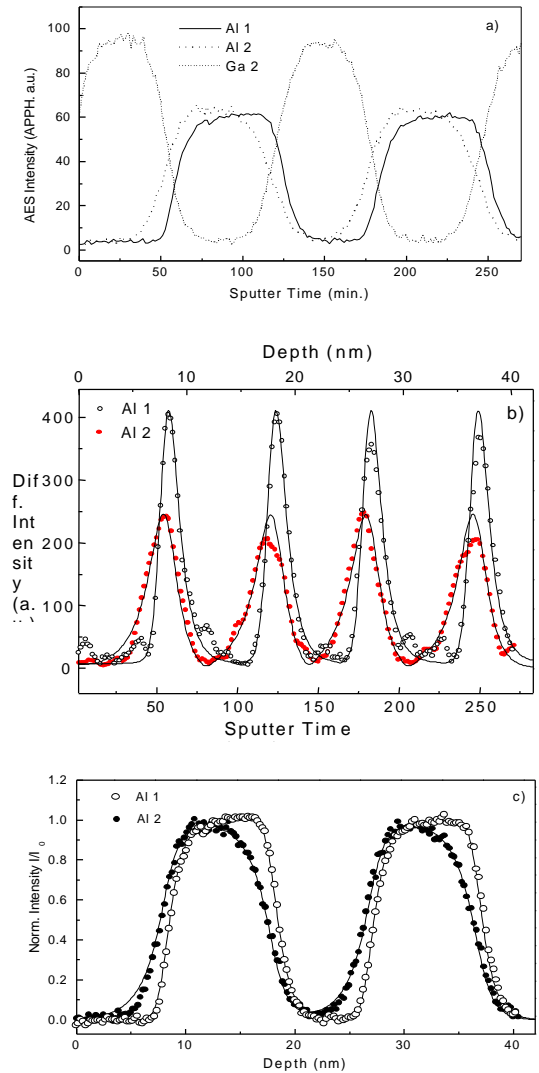

**Fig. 2**: **a)** Sputter depth profile of the first 4 layers of a GaAs/AlAs (8.8/9.9 nm) multilayer using 0.6 keV Ar+ ions and 80 degr. incidence angle. The low (Al1, 68 eV) and high (Al2, 1396 eV) energy AES signal intensities are shown as a function of the sputtering time [11].

**b)** Depth resolution functions (DRFs) calculated by the MRI model with the parameters w = 1.0 nm, σ = 0.6 nm, λ(Al1) = 0.4 nm and λ(Al2) = 1.7 nm

**c)** Profile simulation of a) with the parameters in b) assuming a rectangular distribution of Al with depth.

An example of the application of the MRI model to experimentally determine the DRF is shown in **Fig. 2**. Fig 2a shows the AES sputter depth profile of the first 4 layers of a GaAs/AlAs (8.8/9.9nm) multilayer with atomically flat interfaces, performed with Ar+ ions of 600 eV energy at 80° incidence angle. Both the low (Al1, 68 eV) and high energy (Al2, 1396 eV) Auger peaks of aluminium are used for the Al profile reconstruction [11,47]. The DRFs shown in Fig. 2b (open points for Al2, solid for Al1) were derived with eqn. (3) from the corresponding measured profiles in Fig. 2a, and calculated with the MRI model (solid lines) until optimum fit was obtained with the following parameters: mixing length w= 1.0 nm, roughness $\sigma$ = 0.6 nm, and information depth ( here: Auger electron escape depth) $\lambda(Al1)$ = 0.4 nm, $\lambda(Al2)$ = 1.7 nm. Note that both DRFs are asymmetric and different from a simple Gaussian function. Fig. 2c shows the result of the measured profile simulation with these parameters assuming a sharp rectangular concentration distribution of Al in the AlAs layer sandwiched between two adjacent GaAs layers [11, 47]. As seen by the marked difference of the DRFs for the low and high energy Al peak, for which w and $\sigma$ are the same and only $\lambda$ differs by 1.3 nm (i. e. about 4.5 monolayers), monolayer accuracy is achieved in comparison of different original concentration - depth distributions. The high precision is recognized in the representation of the more rounded top of the AlAs layer profile for the high $\lambda$ (Al2) as compared to the more edge like shape of that for the low $\lambda$ (Al1) measurement. The different shifts of the Al1 and Al2 DRFs and layer profiles with respect to the original interfaces depend on the $\lambda$ and w values. These relations can be used to determine attenuation length values or to establish the depth scale in the respective profile, as shown in another paper [47].

Because of noise in the measured data (e.g. Fig. 2 a)), the experimental DRFs obtained by their differentiation are somewhat corrugated, and smoothing may cause some additional (artificial) broadening [47]. In Fig. 2 b), a slight deterioration with depth probably due to inhomogeneous mixing and increasing roughness is recognized. In any case, measured DRFs on well defined samples with sharp interfaces describe the real situation. As shown here and also by Kitada et al. [55], the MRI model is capable to reproduce measured DRFs sufficiently well for accurate profile reconstruction (e.g. Fig.2 c)) with the additional advantage to get an analytical description based on physically meaningful parameters. Therefore, appropriate DRFs can be predicted for experiments with changed parameters (e. g. $\lambda$-values as in Fig. 2 [11,47]). Even first principle predictons are useful, because estimated values of $\lambda$ and of the mixing length w can be taken from existing data bases for electron attenuation length and for ion ranges, respectively.

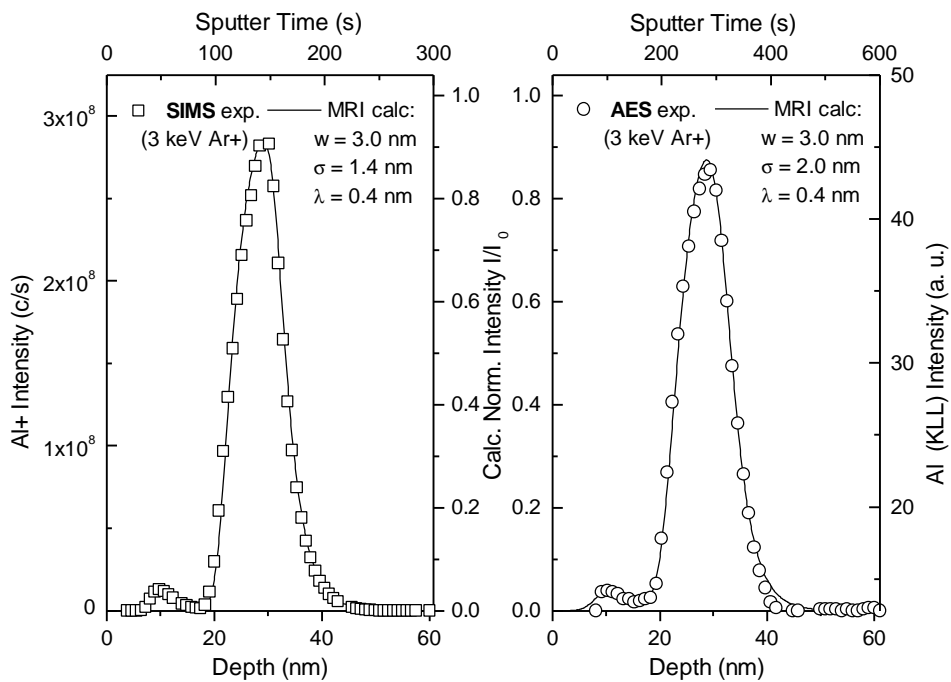

**Fig. 3**

**Fig 3**: Depth profile of a GaAs(10.1 nm)/AlAs(0.28nm)/GaAs(12.8 nm)/AlAs(10.1 nm)/GaAs… sample obtained with SIMS (3keV Ar$^+$, 52 deg inc. angle)(Al$^+$ signal intensity) and AES (3 keV Ar$^+$, 58 deg. inc. angle)(Al LVV peak area), both quantified with MRI calculations with the parameters indicated in the inset [60].

In SIMS profiling, the information depth is given by the escape depth of the secondary ions, which is usually assumed to be 1-2 monolayers and has only a very weak and practically negligible dependence on the primary ion energy [56]. Therefore, the MRI parameter $\lambda$ in SIMS is generally taken as 0.3-0.4 nm (see Fig. 3) and can therefore often be neglected, as demonstrated in one of the first succesful applications of the MRI model to SIMS profiles [57]. Another SIMS profile of $SiO_2$ layers with different thicknesses (0.6-3.5 nm) between $Ta_2O_5$ layers, reported by Kim and Moon [58], gave a precise reconstruction of the original in depth distribution with w=1.6 nm, $\sigma$= 1.0 nm and λ= 0.4 nm, an accuracy of < 0.2 nm and detection of about $1*10^{-3}$ of the corresponding $Si^+$ signal for 100% $SiO_2$ as a constant background [44,47]. A recent comparison between a SIMS ($Al^+$) and an AES (Al LVV) depth profile acquired under comparable experimental conditions is shown in **Fig 3** for a GaAs/AlAs multilayer structure [54, 59] consisting of a delta layer (monolayer=ML) of AlAs beneath 11.5 nm of GaAs, followed by 12.1 nm of another GaAs layer and 36 ML (10.1 nm, 1ML=0.28 nm) of AlAs, sputter depth profiled with 3keV Ar+ ions and 52 deg (SIMS) and 58 deg. (AES) incidence angle, with the MRI parameters (shown in the figures in fairly good agreement [60]. Even the $Al2^+$ cluster ion SIMS profile could be quantified by SIMS assuming an exponential relation between concentration and SIMS intensity [59].

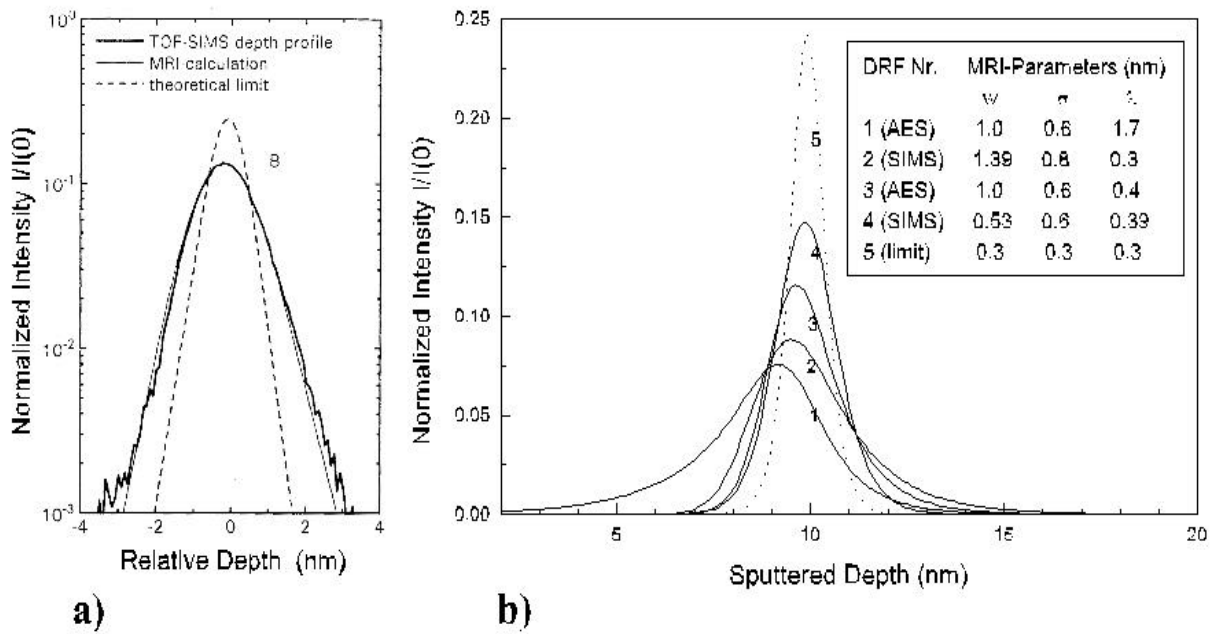

## Fig. 4

**Fig. 4**: **(a)** ToF-SIMS depth profile of a B delta layer in Si after Iltgen et al.[39], fitted with the MRI calculation with the parameters w=0.53 nm, σ= 0.6 nm, λ= 0.39 nm (1ML = 0.28 nm). Courtesy of K. Iltgen [61].(The two lines are identical with lines 4 and 5 on linear scale in (b)).
**(b)** Experimental DRFs fitted with MRI-calculations with the parameters shown in the inset, as compared to the theoretical limit (dashed line). The numbers refer to the following references: 1,3 [11], 2 [57], 4 [39].

As an illustration to experimental DRF determination, **Fig. 4a** shows the SIMS depth profile of a B delta layer in Si after Iltgen et al.[61] and its evaluation by MRI calculation [62] and F**ig 4b** a compilation of some sucessfully applied, typical DRFs for SIMS and AES . In addition, Figs 4 a and b depict the DRF that refers to the theoretical limit, predicted to be about 1 ML for each of the MRI-parameters [52].

**5. Optimization of the experimental depth resolution function**

Determination of the experimental DRF means that the behavior of mixing length, roughness and information depth can be studied as a function of the profiling parameters: ion species, energy and incidence angle in order to find the optimum DRF. Recently this was done using the following GaAs/AlAs structure: (layer thickness in atomic monolayers (ML))[53]:

48GaAs/1AlAs/48GaAs/4AlAs/46GaAs//20AlAs/GaAs(bulk).

AES depth profiling was performed by sputtering with ionized Ar, Xe, and $SF_6$ in the energy range between 500 eV and 1keV at incidence angles between 58 and 80 deg. For reasons of good signal to noise the Ga LMM peak area depth profile was analyzed by means of the MRI model in order to extract roughness and mixing length. The values of the information depth were taken from refs. [52,53].

The MRI parameters were extracted by DRF fitting and are shown in **Fig. 5** as function of the sputtering gas, the ion energy and the ion incidence angle, as described in detail in ref. [63]. In brief, the depth resolution improves in the sequence Ar, Xe, $SF_6$, with decreasing primary ion energy and with increasing incidence angle, as demonstrated in Fig. 5. The depth resolution was calculated from the MRI parameters [8,40].:

$$\Delta z = \left(\Delta z_w^2 + \Delta z_l^2 + \Delta z_s^2\right)^{1/2} = \left[(1.668w)^2 + (1.668l)^2 + (2s)^2\right]^{1/2} \quad (4)$$

The results of independent measurements of w values with ARAES are shown in **Fig. 6.** The physical nature of the w value is confirmed by the good agreement between the value of the mixing length obtained from MRI calculations and the fit of the ARAES

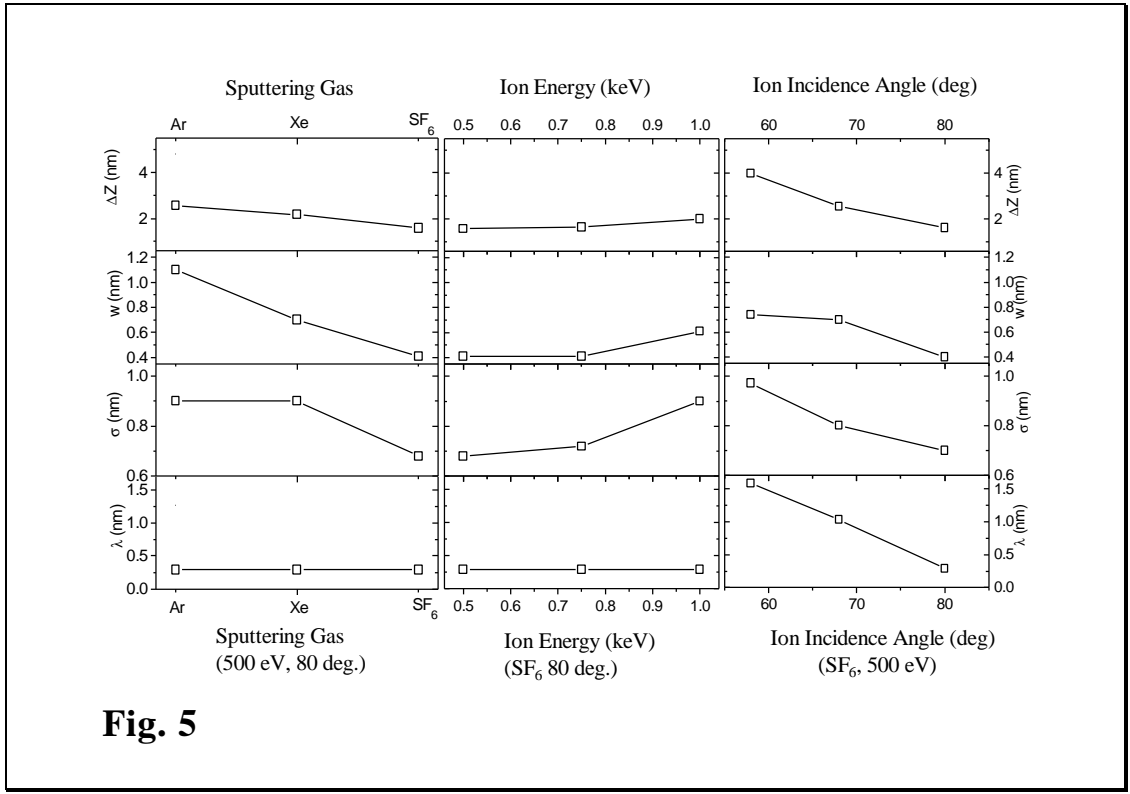

**Fig. 5** Selected dependencies of the MRI parameters mixing length, w, roughness, σ, information depth, λ, and calculated depth resolution Δz (eqn.4) versus type of sputtering gas, ion energy, and ion incidence angle, taken from the AES depth profiles of a 4 ML structure of AlAs in GaAs. For details see refs. [54,63].

measurements with the calculated angle dependence of the ratio of the Al to As intensities [54].

As demonstrated by Iltgen et al. [39] in SIMS profiling, $SF_6$ is particularly useful as a sputtering gas for achievement of ultrahigh depth resolution. As seen in Fig.5, the mixing length w is about 3 times smaller than that of Ar at the same energy ( 500 eV). The explanation for this fact is the decomposition of the impinging ion on impact and the subsequent distribution of the total ionized molecule energy among the atoms [64]. The main primary ion species in ionized $SF_6$ is $SF_5^+$ (total mass 127 amu). Therefore, the sulfur atom (32 amu) gets roughly 25% (=125 eV) and each fluorine atom (19 amu) gets 15% (=75 eV) of the total energy of 500 eV. At these low energies and at glancing icidence, the ion ranges according to TRIM results are indeed abot 0.4-0.6 nm [44], in accordance with MRI calculations and with ARAES measurements.

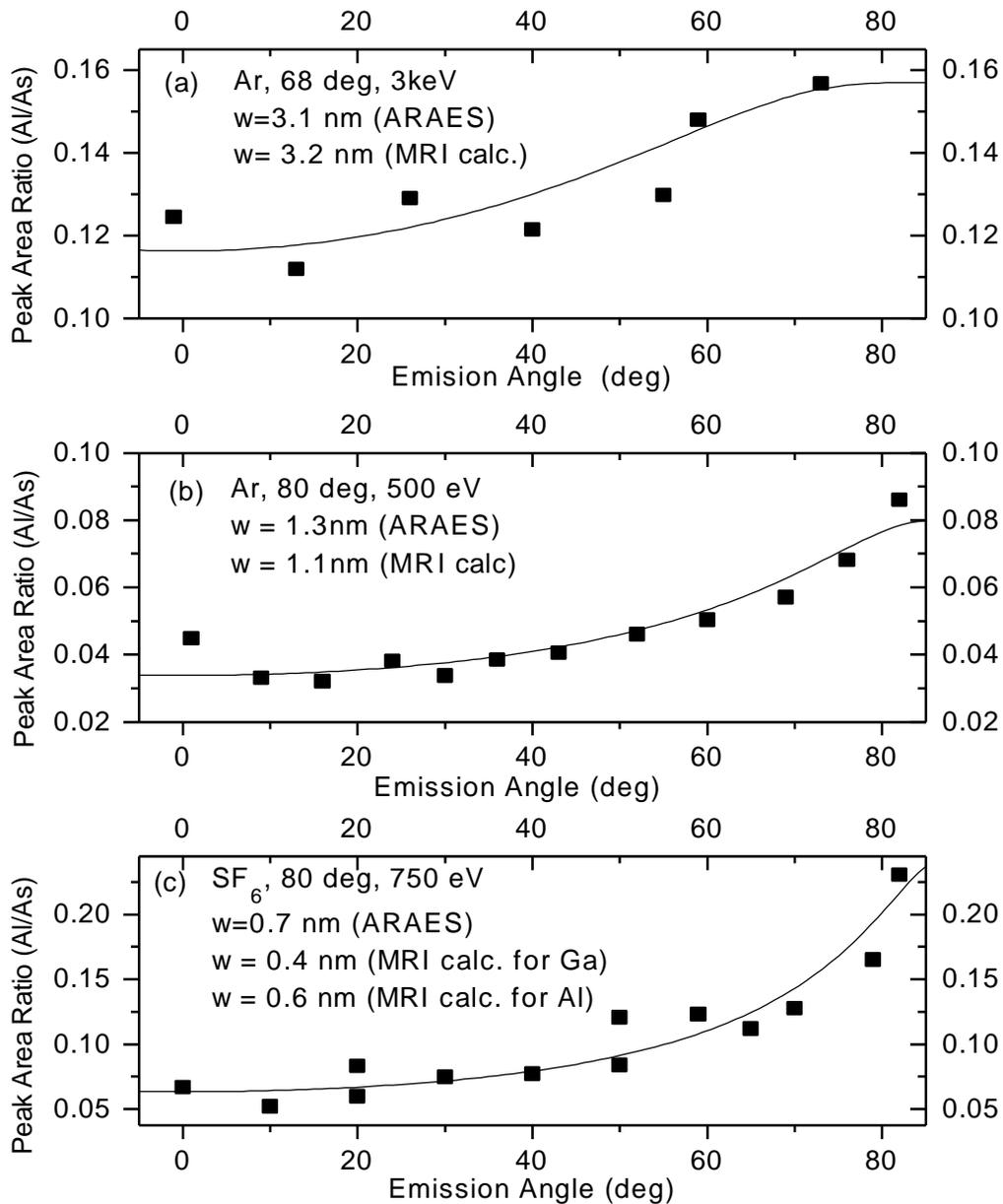

**Fig. 6** Ratio of Al KLL / As LMM peak areas, I(Al)/I(As), versus the emission angle of the Auger electrons, φ, for the AlAs layer structure described in refs. [54,63], to determine the mixing length, w, from the relation I(Al)/I(As)=const.[1-exp[-w/(λcosφ)]], as described in the text, with the electron attenuation length λ=2.32 nm [52,53]. The sputtering conditions are shown in the inset together with the mixing length obtained from the fit with MRI calculations. After ref. [54].

As pointed out by Zalm and Beckers [54] the sputtering yield is the sum of that of each atom in the cluster, and therefore it is not surprising that a similar sputtering yield than with 500 eV Ar+ was obtained.

When the rms roughness from AFM measurements after profiling is compared with the MRI σ parameter, the latter is found systematically higher than the former. This is due to the fact, that the MRI roughness parameter not only contains surface roughness but also includes the straggling of the mixing length, analogous to the ion range straggling [11]. .As an example, **Table1** shows typical values from some of the experiments.

**Tab. 1** Selected values of σ ( = rms roughness) obtained by means of AFM measurements after profiling and by MRI modeling of AES profiles for different typical sputtering conditions. (After ref. [54]).

| Sputtering Gas | energy keV | Angle deg | σ AFM nm | σ MRI nm |
|---|---|---|---|---|
| $SF_6$ | 0.5 | 80 | 0.45 | 0.7 |
| $SF_6$ | 0.5 | 68 | 0.3 | 0.8 |
| Ar | 1 | 80 | 0.3 | 1.2 |
| Original | Surface | | 0.2 | |

As an example, **Table1** shows some selected values from some of the experiments. The order of magnitude of the difference is in agreement with TRIM calculations [33] of ion ranges and range straggling, although for the low energies the TRIM code is not expected to give a similar accuracy as for higher energies. Therefore the MRI roughness parameter σ in the experimental DRF is expected to be:

$$\sigma = \left(\sigma_s^2 + \sigma_w^2 + \sigma_i^2\right)^{1/2} \qquad (5)$$

with $\sigma_i$ the original interface roughness (negligible for high quality reference material),

$\sigma_s$ the rms value of the surface roughness ($=\sigma$ (AFM)) and $\sigma_w$ the mixing length straggling. Eqn (5) shows that $\sigma$ (MRI) is always expected to be larger than $\sigma$ (AFM) in agreement with **Table 1**. For example, the TRIM values for Ar ions of 1keV at 80 deg incidence angle are 1.2 and 0.9 nm for mean range and range straggling, respectively. Eqn. (3) with $\sigma_s = 0.3$ and $\sigma_w = 0.9$ nm (and $\sigma_i = 0$) gives for $\sigma = 0.95$ which is still less than the value used for the fiting calculation (1.2 nm in Ttable 1) but much closer than $\sigma_s$. For sulfur in $SF_6$, with 125 eV impinging energy, the respective values are 0.4 and 0.4 nm for mean range and straggling, respectively. From eqn.(3) it follows $\sigma = 0.57$ nm, again somwhat too small. However, it should be kept in mind that short range roughness measurements with AFM tend to give too low values [63], and that an additional term to mixing range straggling can be attributed to the presence of other lower mass ions such as $SF^+$ and $F^+$ species [39]. Nevertheles, the deviations are below one atomic monolayer.

Recently it was demonstrated that roughness plays a more decisive role for the physical limit of the profile resolving capacity than atomic mixing [50]. For example, when two extreme cases of a DRF are considered, as depicted in **Fig. 7a**, one with a Gaussian shape (roughness) and one with an exponential shape (mixing) and both yield the same $\Delta z$ (in the 16-84% definition at an interface[50]), the resolution capability in terms of distinction of two adjacent monolayers is considerably different. This is shown in **Fig. 7b** for the sputter depth profile of a multilayer consisting of 10 alternating layers with d=5 nm single layer thickness. According to eqn. (4), $\Delta z/d = 2$ is the same for both cases, and we expect the same resolution at a single interface. In contrast to this, the result is a vanishing multilayer profile in case of the Gaussian DRF [1], but a still well resolved multilayer profile in case of the exponential DRF, as shown in Fig.7b [50]. However, the full widths at half maximum (FWHMs) of the DRFs are different (11.8 nm for A and 4.5 nm for B), suggesting the FWHM of the DRF to be a better characterization for the depth resolution in practical cases than the present $\Delta z$ (16-84%) definition [50]. It is evident that reduction of roughness terms, and in particular the "straggling" of the mixing length is most important to obtain ultra high depth resolution.

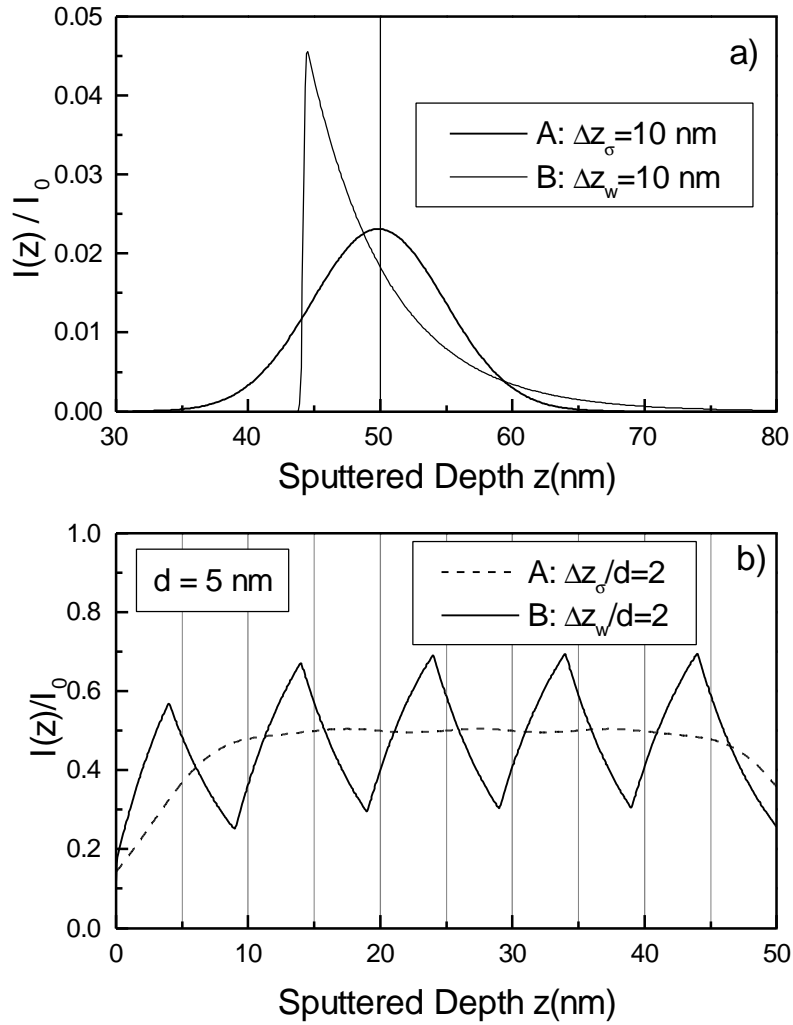

**Fig.7**

**Fig. 7** : **a)** Two typical DRFs (A),(B) (delta layer at z = 50 nm) with the same depth resolution, $\Delta z \approx 10$ nm, but different MRI parameters (dominant σ (A) :$\Delta z \approx \Delta z_\sigma$ and dominant w (B): $\Delta z \approx \Delta z_w$), according to(eqn. 4):
(A): w = 0.1 nm, σ = 5 nm, λ = 0,3 nm  (FWHM =  11.8 nm)
(B): w = 6.0 nm, σ = 0.1 nm, λ = 0.3 nm  (FWHM = 4.5 nm)
**b)** Corresponding profiles of multilayers with 5 nm single layer thickness (i.e. $\Delta z/d \approx 2$ in both cases), determined by MRI profile calculations with DRFs (A) and (B) from Fig. 7a. Note the much higher capability in resolving the multilayer structure of (B) with respect to (A) despite the same $\Delta z$. After ref. [50].

## 6. Conclusions

Although it was recognized a long time ago that appropriate depth resolution functions (DRFs) are necessary for the quantification of depth profiles, i. e. the transformation of sputter depth profiles into depth distributions of composition, only recently the regime of ultra high depth resolution was entered, where the correct shape of the DRF is decisive for the required high accuracy. In order to achieve optimum depth resolution, numerous parameters have to be optimized. Their fundamental limit is given by the minimum values of atomic mixing length, roughness and information depth. Provided there is complete mixing, no additional diffusion and negligible preferential sputtering and segregation, the depth resolution function is well represented by the three respective functional dependencies, two exponentials (mixing, information depth) and one Gaussian (roughness). According to their physical influence on the depth resolution, they are mathematically combined in the so called MRI model describing the DRF. Experimentally, the width of the atomic mixing zone is minimized by using low primary ion energy (typically below 500 eV), and/or the use of molecular ions, and a high incidence angle (typically 80 deg.). While surface roughness is minimized by using sample rotation, mixing length straggling is a more fundamental limitation. The information depth depends on the analysis method. In SIMS, it is determined by the secondary ion escape depth (typically 1-2 monolayers (ML)), whereas in AES and XPS it is given by the electron escape depth (typically 1-10 ML) which depends on the kinetic energy and on the emission angle (Note that considering these possibilities, the information depth of AES or XPS could be made smaller than that of SIMS). The physical limit of any of the MRI parameters seems to be about 1 ML, and their random superposition yields about 3 ML, i.e. the ultimate depth resolution limit appears to be $\Delta z(min) = 0.7…1.0$ nm. Because the resolution power in terms of separation capability of two adjacent features strongly depends on the shape of the DRF, its FWHM is a more adequate characteristic of that capability than the usual $\Delta z(16\text{-}84\%)$. If the depth resolution function is known with high accuracy, e. g. of one monolayer or better, features of this size can be resolved by profile reconstruction using the MRI model. This accuracy, however, requires high quality reference samples with atomically sharp interfaces for the experimental determination of the DRF. Another problem is nonlinear behaviour of the intensity/concentration [59] and of the time/depth scale conversion [57] with respect to sample composition. While principally these effects can be taken into account [57,59], most severe limitations in profile evaluation are encountered when preferential sputtering in conjunction with segregation and radiation enhanced diffusion occur.